\documentclass[12pt]{article}
\usepackage{graphicx} 

\usepackage[small]{titlesec} 
\usepackage[utf8]{inputenc} 
\usepackage[T1]{fontenc}
\usepackage[english]{babel}
\usepackage[left=2cm,right=2cm,top=2cm,bottom=2cm,bindingoffset=0cm]{geometry}

\graphicspath{{figures/}}

\usepackage{setspace}
\usepackage{subcaption}
\usepackage{youngtab}
\usepackage{epigraph}
\usepackage{upgreek}

\usepackage[labelfont=bf, margin=20pt,font={small}]{caption}

\usepackage{amsmath}
\usepackage{esint}
\usepackage{bm}
\usepackage{
    tikz,
	amsmath,
	amssymb,
	amsfonts,
	amsthm,
	amscd,
	comment,
	epsfig,
	color,
	setspace,
	bbold,
	verbatim,
	slashed
}
\usepackage{upgreek}
\usepackage{tikz}
\usetikzlibrary{arrows}
\usepackage{mathrsfs}
\usepackage{multirow}
\usepackage{simpler-wick}
\usepackage{cite}

\newcommand\beq{\begin{equation}}
\newcommand\eeq{\end{equation}}

\tikzset{cross/.style={cross out, draw=black, minimum size=2*(#1-\pgflinewidth), inner sep=0pt, outer sep=0pt},
	cross/.default={5pt}}
\usepackage{mathtools}
\usepackage{graphicx}
\usepackage{authblk}
\usepackage{bbold}
\usepackage[makeroom]{cancel}
\usepackage[hidelinks]{hyperref}

	\definecolor{linkcolor}{rgb}{0,0,1} 
	\definecolor{urlcolor}{rgb}{0,0,1} 
 
\hypersetup{pdfstartview=FitH, linkcolor=linkcolor,urlcolor=urlcolor,citecolor=urlcolor, colorlinks=true}

\usepackage{url}
\usepackage{graphicx}
\usepackage{amssymb}

\usepackage[utf8]{inputenc}
\usepackage{amsmath}
\usepackage{enumitem}



\title{On radiation from hyperbolic motion, behavior of electromagnetic fields, and coordinate transformations at infinity}
\author[1,2]{E.~T.~Akhmedov}
\author[1]{M.~N.~Milovanova}
\affil[1]{\itshape Institutskii per, 9, Moscow Institute of Physics and Technology, 141700, Dolgoprudny, Russia}
\affil[2]{\itshape Academician Kurchatov Square, 1, NRC ''Kurchatov Institute'', 123182, Moscow, Russia}

\begin{document}

{\let\newpage\relax\maketitle}

\begin{abstract}
We show explicitly that radiation from a uniformly accelerating charge escapes outside Rindler wedge, while within Rindler wedge there is no flux through infinity, neither in the Minkowski frame nor in the Rindler frame. This remains true despite the fact that the coordinate transformation between the Rindler and Minkowski frames is not trivial at infinity.
\end{abstract}

\vspace{10mm}
{\bf 1.} The problem of radiation from a uniformly accelerating charge has a long history, dating back to the early works of Max Born \cite{Born} and continuing through numerous investigations in classical and quantum field theory (see, e.g., \cite{Pauli}, \cite{Fulton}, \cite{Peierls} and \cite{Boulware}). A particularly intriguing aspect of this problem is the apparent tension between descriptions of the same physical system in different coordinate frames: while in Minkowski coordinates the field of a uniformly accelerating charge possesses a radiative component, in Rindler coordinates the field appears static and does not radiate. This observation is closely related to the presence of the Rindler horizon and has been widely discussed in the literature (see, for example, \cite{Unruh:1976db}, \cite{Wald}, \cite{Weinberg}).

In our previous work \cite{Akhmedov:2025} we revisited this problem and demonstrated explicitly that in Rindler coordinates a uniformly accelerating charge does not radiate, since this region constitutes a static zone for the source. In contrast, in Minkowski coordinates there exists a wave zone in which the same charge exhibits radiative behavior. This apparent discrepancy calls for a more detailed and quantitative comparison of the two descriptions, which we undertake in the present work.

A key geometric feature underlying this difference is that the Rindler metric cannot be globally represented as a pullback of the Minkowski metric by a smooth, non-degenerate coordinate transformation. Although locally the two metrics are related by a coordinate change, the corresponding Jacobian becomes singular at the Rindler horizon. As a consequence, the mapping between the two coordinate systems is not globally invertible, and a portion of the electromagnetic field effectively propagates beyond the Rindler horizon, remaining inaccessible to non-inertial co-accelerating observers. This feature is intimately connected with the well-known causal structure of Rindler coordinates and the role of horizons in field theory (cf. \cite{Unruh:1976db}, \cite{Fulling}).

The same geometric subtlety manifests itself in the analysis of the asymptotic behavior of the fields. In particular, the asymptotics obtained directly in Minkowski coordinates may, in principle, differ from those reconstructed via Rindler coordinates using the Jacobian transformation. This is because the transformation becomes singular in the relevant limits, and therefore the operations of taking asymptotics and performing coordinate transformations do not necessarily commute. Closely related issues arise in the general theory of the asymptotic structure of fields in spacetime, as developed in the seminal works \cite{Penrose} and \cite{Bondi}, where the behavior of fields at infinity is known to be sensitive to the choice of coordinates and conformal compactification.

The purpose of the present paper is to clarify these issues in the specific context of electromagnetic fields produced by a uniformly accelerating charge. We perform a detailed comparison between two procedures for obtaining asymptotic behavior: (i) a direct analysis in Minkowski coordinates, and (ii) an indirect approach based on asymptotics in Rindler coordinates followed by a transformation using the Jacobian matrix. 

{\bf 2.} We consider Maxwell's equations in curvilinear coordinates,
\begin{align} \label{eom}
\frac{1}{\sqrt{-g}} \partial_{\mu} \left(\sqrt{-g} \, F^{\mu \nu}\right) = 4 \pi j^{\nu},
\end{align}
where $g_{\mu\nu}$ is the metric tensor, $g=\det g_{\mu\nu}$, $F^{\mu\nu}=\partial^\mu A^\nu - \partial^\nu A^\mu$ is the electromagnetic field tensor, $A_\mu$ is the four-potential, and $j^\nu$ is the current density. This current is due to a uniformly accelerated charge with constant proper acceleration $a$, moving along the worldline
\begin{align} \label{worldline}
z^{\mu}(\theta) = \left( \frac{1}{a} \sinh(a\theta), \frac{1}{a} \cosh(a\theta), 0, 0 \right),
\end{align}
where $\theta$ is the proper time.

We consider these equations in two coordinate systems covering flat spacetime: Minkowski coordinates,
\begin{align} \label{minkowski_metric}
ds^2 = dt^2 - dx^2 - dy^2 - dz^2,
\end{align}
and Rindler coordinates,
\begin{align} \label{rindler_metric}
ds^2 = \rho^2 d\tau^2 - d\rho^2 - dy^2 - dz^2,
\end{align}
which cover only the Rindler wedge $|t|<x$ of the entire flat spacetime, provided $\rho \geq 0$.

The coordinate transformation between these systems,
\begin{align} \label{coordinate_transformation}
t = \rho \sinh \tau, \qquad x = \rho \cosh \tau,
\end{align}
(with $y$ and $z$ unchanged) does not reduce to a trivial transformation at infinities ($t\to\pm \infty$, $x\to \pm \infty$) and, thus, is not quite a proper gauge transformation. Such a transformation can modify the asymptotic behavior of the fields and, consequently, the associated stress–energy flux at infinity. The purpose of this work is to analyze this possibility in detail and to clarify in what sense the homogeneously accelerating charge does create electromagnetic radiation, whereas in the Rindler wedge the radiation is absent \cite{Pauli,Kalinov,Akhmedov:2025}.

To analyze this situation, one must first identify the relevant asymptotic regions. In both descriptions, the electromagnetic field is nonvanishing only at spacetime points satisfying the matching (retardation) condition, which enforces a lightlike separation between the emission and observation points. In Minkowski coordinates, this condition takes the form
\begin{align} \label{connection_minkowski}
\left(t-\frac{\sinh (a \theta )}{a}\right)^2 = \left(x - \frac{\cosh (a \theta )}{a}\right)^2+y^2+z^2.
\end{align}
Here $\theta$ is the proper time of emission and $(t,x,y,z)$ is the observation point. We set the speed of light to one. Applying the transformation \eqref{coordinate_transformation}, one obtains the corresponding relation in Rindler coordinates:
\begin{align} \label{connection_rindler}
    \cosh (\tau -a \theta ) = \frac{a}{2 \rho} \left(\rho^2 + y^2 + z^2 + a^{-2} \right),
\end{align}
where $(\tau, \rho, y, z)$ is the observation point within the Rindler wedge.

{\bf 3.} Analysis of these relations within the Rindler wedge reveals two distinct asymptotic regimes. In Rindler coordinates, for $\tau \to \pm \infty$,
\begin{enumerate}[label=\text{R\arabic*.}, ref=R\arabic*]
\item
$\rho \approx e^{\pm (\tau - a \theta)} \left(1 - \frac{1 + a^2 (y^2 + z^2) e^{\pm 2 a \theta}}{e^{\pm 2 \tau}} \right),
\quad y, z \ll \rho,$
\label{item_rindler_1}
\item
$y^2+z^2 \approx a^{-1} \rho e^{\pm (\tau - a \theta)} \left( \frac{1}{2} \pm \frac{\sinh(a \theta) e^{\pm a \theta}}{e^{\pm 2 \tau}} \right),
\quad \rho^2 \ll y^2+z^2, \quad \rho \not\to 0.$
\label{item_rindler_2}
\end{enumerate}

The corresponding regimes in Minkowski coordinates, for $t \to \pm \infty$, are
\begin{enumerate}[label=\text{M\arabic*.}, ref=M\arabic*]
\item
$t \approx \pm x \mp a^{-1} e^{\mp a \theta}$,
\quad $y^2 + z^2 \ll a^{-1} x,$
\label{item_minkowski_1}
\item
$t \approx \pm x \mp a^{-1} e^{\mp a \theta} \pm \frac{y^2 + z^2}{x}$,
\quad $y^2 + z^2 \sim a^{-1} x.$
\label{item_minkowski_2}
\end{enumerate}
These limits are in one-to-one correspondence: \ref{item_rindler_1} with \ref{item_minkowski_1}, and \ref{item_rindler_2} with \ref{item_minkowski_2}. Note that both sets of events $(\tau, \rho, y, z)$ and $(t, x, y, z)$ lie in the same Rindler wedge.

\begin{figure}[h]
    \centering
    \includegraphics[width=0.4\textwidth]{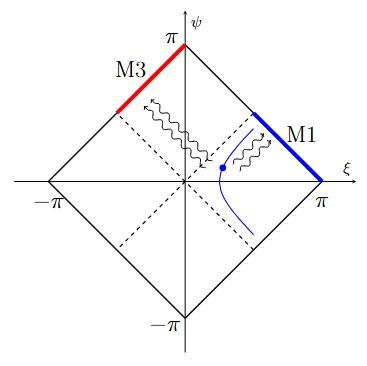}
    \caption{The Penrose diagram of Minkowski spacetime illustrating the asymptotic regions M1 and M3, together with the worldline of the charge, provides a useful geometric interpretation. The wavy lines represent tentative electromagnetic radiation propagating along the future light cone, extending into regions that lie beyond the Rindler wedge.}
    \label{figure:penrose}
\end{figure}

We now illustrate the comparison using the first regime, \ref{item_rindler_1} and \ref{item_minkowski_1}, which corresponds to the right future lightlike infinity $\mathcal{J}^{+}$ on the Penrose diagram, with small values of the transverse coordinates $y$ and $z$ (see Fig.~\ref{figure:penrose}). The second regime, \ref{item_rindler_2} and \ref{item_minkowski_2}, corresponds instead to the asymptotic region characterized by large transverse coordinates $y$ and $z$.

In Rindler coordinates, the magnetic field vanishes, while the electric field is nonzero \cite{Akhmedov:2025}, \cite{Kalinov}:
\begin{align} \label{e_rindler}
\vec{E}^{R} = \frac{1}{\rho \sinh^3 (\tau -a \theta )}
\begin{pmatrix}
a \rho - \cosh (\tau -a \theta ) \\
a y \\
a z
\end{pmatrix}, \quad \vec{B}^{R} = 0.
\end{align}
Here the superscript $R$ means that the fields are defined in the Rindler corrdinates.
In the regime \ref{item_rindler_1}, corresponding to $\rho \to \infty$, the leading asymptotics are
\begin{align} \label{e_limit_rindler}
\vec{E}^{R} \simeq \pm 4 a e^{\mp 3 (\tau - a \theta)}
\begin{pmatrix}
1 \\
2 a y e^{\mp (\tau - a \theta)} \\
2 a z e^{\mp (\tau - a \theta)}
\end{pmatrix}, \quad \vec{B}^{R} = 0.
\end{align}
The transformation of the electromagnetic tensor between the two coordinate systems is given by
\begin{align}
F_{\mu \nu}^{M} = J^{\sigma}_{\mu} F_{\sigma \rho}^{R} J^{\rho}_{\nu},
\end{align}
with the Jacobian equal to
\begin{align} \label{jacobi_matrix}
J^{\mu}_{\nu} =
\begin{pmatrix}
\frac{\cosh(\tau)}{\rho} & \frac{-\sinh(\tau)}{\rho} & 0 & 0 \\
-\sinh(\tau) & \cosh(\tau) & 0 & 0 \\
0 & 0 & 1 & 0 \\
0 & 0 & 0 & 1
\end{pmatrix}, 
\end{align}
for the two frames under consideration. 

In the limit \ref{item_rindler_1}, the Jacobian behaves asymptotically as
\begin{align} \label{jacobi_matrix_first_variant}
J^{\mu}_{\nu} \simeq
\begin{pmatrix}
\frac{a e^{\pm a \theta}}{2} \left(1 + \frac{1 + a^2 (y^2 + z^2) e^{\pm 2 a \theta}}{e^{\pm 2 \tau}} \right) &
\mp \frac{a e^{\pm a \theta}}{2} \left(1 + \frac{-1 + a^2 (y^2 + z^2) e^{\pm 2 a \theta}}{e^{\pm 2 \tau}} \right) & 0 & 0 \\
\mp \left( \frac{e^{\pm \tau}}{2} - e^{\mp \tau} \right) & e^{\pm \tau} & 0 & 0 \\
0 & 0 & 1 & 0 \\
0 & 0 & 0 & 1
\end{pmatrix}.
\end{align}
Combining Eqs.~\eqref{e_limit_rindler}--\eqref{jacobi_matrix_first_variant}, one finds
\begin{align}
\vec{E}^{M} \simeq \pm 4 a^2 e^{\mp 4 (\tau - a \theta)}
\begin{pmatrix}
1 \\
a y e^{\pm a \theta} \\
a z e^{\pm a \theta}
\end{pmatrix},
\quad
\vec{B}^{M} \simeq 4 a^2 e^{\mp 4 (\tau - a \theta)}
\begin{pmatrix}
0 \\
- a z e^{\pm a \theta} \\
a y e^{\pm a \theta}
\end{pmatrix}.
\end{align}
Here the superscript $M$ means that the fields are defined in the Minkowski corrdinates.

Thus, one obtains a nonvanishing magnetic field. Furthermore,
after using the matching relation $e^{\pm \tau} \approx a \rho e^{\pm a \theta}$ valid in this limit and subsequently expressing the result in Minkowski coordinates, one recovers the same asymptotic expressions. Indeed, the electric and magnetic fields in Minkowski coordinates are given by \cite{Akhmedov:2025}, \cite{Kalinov}:
\begin{align} \label{e_b_minkowski}
\vec{E}^{M} &= \frac{1}{(t \cosh(a \theta) - x \sinh (a \theta ))^3}
\begin{pmatrix}
a (x^2-t^2)-x \cosh(a \theta) + t \sinh (a \theta ) \\
a x y \\
a x z
\end{pmatrix}, \\
\vec{B}^{M} &= \frac{1}{(t \cosh(a \theta) - x \sinh (a \theta ))^3}
\begin{pmatrix}
0 \\
- a t z \\
a t y
\end{pmatrix}. \nonumber 
\end{align}
In the limit \ref{item_minkowski_1}, their asymptotic form is
\begin{align} \label{e_b_limit_minkowski}
\vec{E}^{M} &\simeq \pm \frac{e^{\pm 2 a \theta}}{x^2}
\begin{pmatrix}
1 \\
a y e^{\pm a \theta} \\
a z e^{\pm a \theta}
\end{pmatrix}, \qquad
\vec{B}^{M} \simeq \frac{e^{\pm 2 a \theta}}{x^2}
\begin{pmatrix}
0 \\
- a z e^{\pm a \theta} \\
a y e^{\pm a \theta}
\end{pmatrix}.
\end{align}
Note that these fields decay too rapidly at infinity, namely as $1/x^2$. As a consequence, the flux through future lightlike infinity $\mathcal{J}^{+}$ within the Rindler wedge (see Fig.~\ref{figure:penrose}), where the fields are simultaneously described in both coordinate systems, vanishes. Therefore, no radiation is produced within the Rindler wedge, neither in Rindler nor in Minkowski coordinates.

{\bf 4.} We now demonstrate that radiation is indeed present and is emitted into the region beyond the Rindler wedge. This becomes manifest when the problem is analyzed in Minkowski coordinates using the corresponding expressions for the electromagnetic fields. The crucial point is that the expressions (\ref{e_b_minkowski}) remain valid not only inside the Rindler wedge, but also in the complementary regions of Minkowski spacetime lying beyond it.

The asymptotic regime in Minkowski spacetime outside the Rindler wedge, corresponding to the region $x<t$, can be parameterized as
\begin{enumerate}[label=\text{M3.}, ref=M3]
\item
$ t \approx \pm \sqrt{x^2 + y^2 + z^2} + a^{-1} \sinh(a \theta) \mp \frac{2 a^{-1} x \cosh(a \theta)}{\sqrt{x^2 + y^2 + z^2}}$.
\end{enumerate}
In this regime, the leading asymptotic behavior of the electromagnetic fields can likewise be obtained from Eq.~(\ref{e_b_minkowski}) together with the matching condition (\ref{connection_minkowski}) in the limit $t\to \infty$, now evaluated in the region beyond the Rindler wedge:

\begin{align}
\vec{E}^{M} &\simeq \frac{1}{\left( \sqrt{x^2 + y^2 + z^2} \cosh(a \theta) - x \sinh (a \theta ) \right)^3}
\begin{pmatrix}
-a (y^2+z^2)+ 3 x \cosh(a \theta) - \sqrt{x^2 + y^2 + z^2} \sinh (a \theta ) \\
a x y \\
a x z
\end{pmatrix}, \\
\vec{B}^{M} &\simeq \frac{\sqrt{x^2 + y^2 + z^2}}{\left( \sqrt{x^2 + y^2 + z^2} \cosh(a \theta) - x \sinh (a \theta ) \right)^3}
\begin{pmatrix}
0 \\
- a z \\
a y
\end{pmatrix}. \nonumber
\end{align}
To determine whether the charge radiates into this asymptotic region, we compute the flux of electromagnetic energy through a sphere at infinity, denoted by $\Omega$. Let $t = R$, and parametrize the observation point as $x = R \cos(\phi), y = R \sin(\phi) \cos(\chi), z = R \sin(\phi) \sin(\chi)$ with $R \to +\infty$. The corresponding unit normal vector is $\vec{n} = (\cos(\phi), \sin(\phi) \cos(\chi), \sin(\phi) \sin(\chi))$. The energy flux per unit solid angle at infinity is then given by
\begin{align}
    \frac{d I}{d \Omega} = \left( \vec{S}, \vec{n} \right) R^2 \approx \frac{a^2 \left(y^2 + z^2 \right) \left(x^2 + y^2 + z^2 \right)^2}{\left( \sqrt{x^2 + y^2 + z^2} \cosh(a \theta) - x \sinh (a \theta ) \right)^6} = \frac{a^2 \sin^2(\phi)}{\left(\cosh(a \theta) - \cos(\phi) \sinh (a \theta ) \right)^6}. 
\end{align}
Integrating this expression over the angular variables $\phi$ and $\chi$, one obtains a nonvanishing total energy flux. This demonstrates that the homogeneously accelerating charge does radiate; however, the radiation propagates into the asymptotic region lying beyond the Rindler wedge.

\noindent\textbf{Acknowledgments.} This work was supported by grant No. 26-12-00330 from the Russian Science Foundation (RSF).

\end{document}